\documentclass[aip,amsmath,amssymb,preprint, twocollumns]{revtex4-1}	
\usepackage{graphicx}
\usepackage{psfrag}
\usepackage{ae}
\usepackage{amsmath,amssymb}

\begin{document}
\title{Confinement effects on the properties of Janus dimers}

\author{Jos\'e Rafael Bordin}
\email{josebordin@unipampa.edu.br}
\affiliation{Campus Ca{\c{c}}apava do Sul, Universidade Federal do Pampa, Av. Pedro Anuncia\c c\~ao 111, 
CEP 96570-000, Ca\c{c}apava do Sul, Brazil. E-mail: josebordin@unipampa.edu.br}

\author{Leandro B. Krott}
\affiliation{Centro Ararangu\'a, Universidade Federal 
de Santa Catarina, Rua Pedro Jo\~ao Pereira, 150, 
CEP 88905-120, Ararangu\'a, SC, Brazil. E-mail: leandro.krott@ufsc.br}

\begin{abstract}

Confinement has been suggested as a tool to tune the self-assembly 
properties of nanoparticles, surfactants, polymers and colloids.
In this way, we explore the phase diagram of Janus nanoparticles using Molecular Dynamics simulations.
The nanoparticle was modeled as a dimer made by one monomer
that interacts by a standard Lennard Jones potential and another monomer that
is modeled using a two-length scale shoulder potential. This specific design
of nanoparticle exhibits in bulk distinct self-assembled structures and 
water-like diffusion anomaly. Our results indicate that 
besides the aggregates observed in bulk, new structures are observed 
under confinement. Also, the dynamic and thermodynamic behavior
of the fluid phase are affected. The systems show a reentrant fluid phase and  
density anomaly. None of these two features were observed in bulk. Our results show that geometrical confinement
leads to new structural, thermodynamical and dynamical behavior for this Janus nanoparticle.

\end{abstract}

\maketitle

\section{Introduction}
\label{introduction}

Colloids and molecules with anisotropic shapes and interactions play a significant role in
condensed matter physics, specially in the design of self-assembled structures~\cite{Roh05, Klapp16}. 
Particularly, Janus colloids are characterized as particles composed by at least two physically or chemically
distinctive surfaces. They can have different shapes, as rods, spheres and dumbbells.
These systems have a large range of applications including medicine, self-driven molecules,
catalysis, photonic crystals, stable emulsions, biomolecules and self-healing materials\cite{Cas89, Talapin10,ElL11, TuP13, WaM08,
WaM13, Zhang15, Bic14, Ao15}.

From the distinct Janus particles shapes,
Janus dumbbells\cite{Yin01, SiC14, Lu02, YoL12} are colloids formed by two spheres,
each one with distinct characteristics
linked together with a separation that varies from an almost total overlap to one or two monomer diameters. 
Due to the resemblance between Janus particles and competing
interaction systems, Janus dumbbells behave
as surfactant in water-based emulsions due its amphiphilic 
properties~\cite{SoK11, TaI05, Liu13}.
Self-assembly lamellae or micellae phases were observed on these systems due the competition between attractive 
and repulsive forces \cite{Li12, White10, Munao13, Munao14, Munao15b, Avvisati14}.

Recent studies reported the production of silver-silicon (Ag-Si)\cite{SiC14} and silica-polystyrene (SiO$_2$-PS)\cite{Liu09}
Janus dimers. Silica and silicon are classified as anomalous fluids, and therefore have a set of properties that
diverge from the observed for regular materials.
For most part of the fluids, 
the diffusion coefficient decreases when the pressure (or density)
increases. However, materials as water~\cite{Ne02a}, silicon~\cite{Mo05} and silica~\cite{Sa03} 
show diffusion anomaly,  characterized by a
maximum in the diffusion coefficient at constant temperature. Besides diffusion (or dynamical) anomaly,
water, silicon, silica and others fluids, the so-called anomalous fluids, also have
other classes of anomalies, as structure and thermodynamic anomalies. Particularly,
the density anomaly is characterized by the increase of density with the temperature 
at a fixed pressure.

Distinct computational models were proposed to study the anomalous behavior of fluids.
Among these models, effective two length scale (TLS) core-softened shoulder potential 
are an interesting tool to investigate systems with water-like anomalies. Particularly, the
model proposed by Oliveira~
\textit{et. al.}\cite{Ol06a, Ol06b} reproduces qualitatively the diffusion, structural and thermodynamic anomalies, 
and was broadly used to study anomalous systems. 
This effective approach to describe anomalous fluids was used to study monomeric and dimeric systems of anomalous
particles~\cite{Ol06a, Ol06b, Ol10, Ga14, Munao16}. The TLS potential was used in our previous works to study the 
behavior of Janus dumbbells composed of one anomalous and one non-anomalous monomers
in bulk~\cite{BoK15c, Bordin16a}. Despite the presence of the non-anomalous monomer, the diffusion anomaly was preserved.

Confinement was proposed as a
approach to tune the self-assembled morphologies. Controlling the confinement
intensity it is possible to create micelles with distinct shapes.
Computational and experimental studies have already explore the confinement
effects if the self assembly of polyhedral nanoparticles~\cite{Khad16}, patchy spherical colloids~\cite{Iwa16},
asymmetric and symmetric dumbbells~\cite{Lee09, Muang14}
surfactants and polymers~\cite{Kim15, Ro12, Ro11}. 
As well, the confinement affects the diffusivity of spherical Janus swimmers~\cite{Ao15}.
In fact, confinement strongly affects the behavior of fluids in general. For the case of
anomalous fluids~\cite{KoB15}, new anomalies can be observed due confinement~\cite{BoK15a},
and even a superdiffusive regime can be induced~\cite{BoK14c}.
The anomalous region in the pressure $\times$ temperature ($PT$) phase diagram
is usually shifted due confinement. This shift can be to higher or lower temperatures, 
regarding on the nature of the fluid-wall interaction~\cite{Krott14}.

Therefore, the question that rises is how the confinement will affect not only the 
self-assembled structures, but the dynamical and thermodynamical behavior
of the anomalous/non-anomalous Janus dumbbell
system. In this way, we perform intensive Molecular Dynamics (MD)
simulations of Janus nanoparticles composed of anomalous and non-anomalous monomers
confined between two flat plates.
In addition to the myriad of structures tuned by the confinement, we show
how the confinement and the TLS potential lead the system to have not only diffusion anomaly, but also
the density anomaly, not observed in bulk.

The paper is organized as follows: first we introduce the model and describe the methods 
and simulation details; next the results and discussion are given; and 
then we present our conclusions.

\section{The Model and the Simulation details}
\label{Model}

In this paper all physical quantities are computed in the standard Lennard Jones (LJ) units\cite{AllenTild},

\begin{equation}
\label{red1}
r^*\equiv \frac{r}{\sigma}\;,\quad \rho^{*}\equiv \rho \sigma^{3}\;, \quad 
\mbox{and}\quad t^* \equiv t\left(\frac{\epsilon}{m\sigma^2}\right)^{1/2}\;,
\end{equation}

\noindent for distance, density of particles and time , respectively, and

\begin{equation}
\label{rad2}
p^*\equiv \frac{p \sigma^{3}}{\epsilon} \quad \mbox{and}\quad 
T^{*}\equiv \frac{k_{B}T}{\epsilon}
\end{equation}

\noindent for the pressure and temperature, respectively, where $\sigma$, $\epsilon$ and $m$ 
are the distance, energy and mass parameters, respectively.
Since all physical quantities are defined in reduced LJ units, 
the $^*$ is  omitted, in order to simplify the discussion.

The systems have $N = 1000$ dimeric particles, totalizing $N = 2000$ particles, confined between two smooth and parallel 
plates. The Janus dumbbells particles were modeled using two spherical core-softened particles, each one with mass $m$ 
and effective diameter $\sigma$, linked rigidly at a distance $\lambda$. The dimers are formed by 
monomers of type A and type B.

The particles of type A present anomalous behavior and their interaction is given by a two length scales potential, 
potential AA, defined as~\cite{Ol06a, Ol06b}

 $$
 \frac{U^{AA}(r_{ij})}{\varepsilon} = 4\left[ \left(\frac{\sigma}{r_{ij}}\right)^{12} -
 \left(\frac{\sigma}{r_{ij}}\right)^6 \right] + 
 $$
 \begin{equation}
 u_0 {\rm{exp}}\left[-\frac{1}{c_0^2}\left(\frac{r_{ij}-r_0}{\sigma}\right)^2\right]\;,
 \label{AlanEq}
 \end{equation}

\noindent where $r_{ij} = |\vec r_i - \vec r_j|$ is the distance between two A particles $i$ and $j$. The first term of the potential
is a standard 12-6 LJ potential~\cite{AllenTild} and the second one is a Gaussian shoulder centered at $r_0$, 
with depth $u_0$ and width $c_0$. The parameters used in this work are $u_0 = 5.0$, $c_0 = 1.0$ and $r_0/\sigma = 0.7$.
Both systems, monomeric and dimeric, modeled by this potential, present density, diffusion and thermodynamic anomalies, 
like observed in water, silica and other anomalous fluids~\cite{Ol06a, Ol06b, Ol10, Kell67,Angell76}.

\begin{figure}[ht]
\begin{center}
\includegraphics[width=6cm]{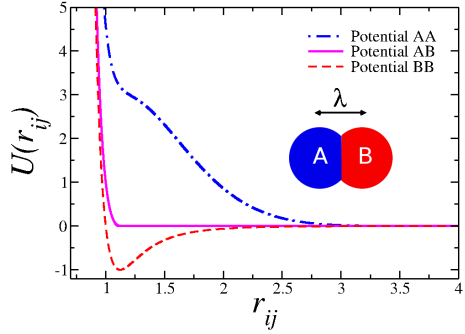}
\end{center}
\caption{Interaction potential between particles of type A (dot-dashed blue line), 
between particle of type A and B (solid magenta line) and between particles of type B (dashed red line).
The interaction between dimers and confining walls is given by the projection of the potential AB
in $z$-direction. Inset: Janus dumbbells formed by A-B monomers.}
\label{fig1}
\end{figure}

The interaction between particles of type B, the potential BB, is given by a standard 12-6  LJ potential, like the first term of
Eq.~\ref{AlanEq}, cut and shifted at the cutoff radius $r_c$,

 \begin{equation}
 \label{LJCS}
 U^{\rm{CSLJ}}(r_{ij}) = \left\{ \begin{array}{ll}
 U_{{\rm {LJ}}}(r_{ij}) - U_{{\rm{LJ}}}(r_c)\;, \qquad r_{ij} \le r_c\;, \\
 0\;, \qquad \qquad \qquad \qquad \quad r_{ij}  > r_c\;.
 \end{array} \right.
 \end{equation}

 The BB potential has a cutoff radius of $r_c = 2.5$. Meanwhile, the interaction for A-B particles
is given by the Weeks-Chandler-Andersen (WCA) potential, defined by the equation~\ref{LJCS}
with $r_c = 2^{1/6}$ is the cutoff. The
interactions between dimers and walls are given by the projection of the WCA potential in the 
$z$-direction. The potentials are illustrated in Fig.~\ref{fig1}.

The simulations were performed in the canonical ensemble using the 
ESPResSo package~\cite{espresso1, espresso2}.
The number density, defined as $\rho = N/V$, where $V=L^2\times L_z$ is the volume of the 
simulation box, was varied from $\rho = 0.05$ to $\rho = 0.50$. 
In all simulations, $L_z=4.0$ and $L$ was obtained from $L = [N/(\rho L_z)]^{1/2}$.
Standard periodic boundary conditions are applied in $x$ and $y$-directions. 
The temperature was simulated in the interval between $T = 0.05$ and $T = 0.60$.
The system temperature was fixed using the Langevin thermostat 
with $\gamma = 1.0$,
and the equations of motion for the fluid particles were integrated
using the velocity Verlet algorithm, with a time step $\delta t = 0.01$. 
We performed $1\times10^6$ steps to equilibrate the system. 
These steps are then followed by $5\times10^6$ steps for the results 
production stage. 
To ensure that the system was equilibrated, the pressure, kinetic energy
and potential energy were analyzed as function of time, as well several 
snapshots at distinct simulation times.
Since confined systems can be sensitive to the number of particles
in the simulation, in some points we carried out simulations with 
5000 and 10000 particles, and 
essentially the same results were observed. As well, 
we run some points with a production time of $1\times10^8$
to test if the system was well equilibrated,
and the same results were obtained.

The system dynamics was analyzed using the lateral mean square displacement (LMSD) as function of time, given by

\begin{equation}
\label{r2}
\langle [\vec r_{\parallel\rm cm}(t) - \vec r_{\parallel\rm cm}(t_0)]^2 \rangle =\langle \Delta \vec r_{\parallel\rm cm}(t)^2 \rangle\;,
\end{equation}

\noindent where $\vec r_{\parallel\rm cm}(t_0) = (x_{\rm cm}(t_0)^2 + y_{\rm cm}(t_0)^2)$ 
and  $\vec r_{\parallel\rm cm}(t) = (x_{\rm cm}(t)^2 + y_{\rm cm}(t)^2)$
denote the parallel coordinate of the nanoparticle center of mass (cm)
at a time $t_0$ and at a later time $t$, respectively. The LMSD is related to the lateral
diffusion coefficient, $D_{\parallel}$, by

\begin{equation}
 D_{\parallel} = \lim_{t \rightarrow \infty} \frac{\langle \Delta \vec r_{\parallel\rm cm}(t)^2 \rangle}{4t}\;.
\end{equation}

The pressure in confined systems by parallel plates is divided in parallel and perpendicular direction.
The parallel pressure, $P_{\parallel}$, was obtained from

$$
P_{\parallel} = 0.5(\sigma_{xx} + \sigma_{yy})\;,
$$
\noindent where $\sigma_{xx}$ and $\sigma_{yy}$ are the normal stress in the $x$ and $y$ direction.

The system structure was analyzed with the lateral radial distribution function (LRDF) $g_{||}(r)$, defined as~\cite{Ku05b}
\begin{equation}
\label{gr_lateral}
g_{||}(r) \equiv \frac{1}{\rho ^2V}
\sum_{i\neq j} \delta (r-r_{ij}) \left [ \theta\left( \left|z_i-z_j\right| + \frac{\delta z}{2}
\right) - \theta\left(\left|z_i-z_j\right|-\frac{\delta z}{2}\right) \right],
\end{equation}
\noindent where $\delta(x)$ is the Dirac $\delta$ function
and the Heaviside function $\theta (x)$ restricts the sum of particle pair in the same
slab of thickness $\delta z = \sigma$. The lateral radial
distribution function is proportional to the probability of finding a particle
at a distance $r$ from a referent particle inside the slab of thickness
$\delta z$.

In order to check if the Janus system shows density anomaly we evaluate the 
temperature of maximum density (TMD). Using thermodynamical relations, the
TMD can be characterized by the minimum of the pressure versus
temperature along isochores,
 \begin{equation}
  \left(\frac{\partial P_{||}}{\partial T}\right)_{\rho} = 0\;.
  \label{TMD}
 \end{equation}Confinement effects on the properties of Janus dimers
\noindent The fluid, micellar and aggregated regions in the $P_{\parallel}  T$ phase diagrams were defined 
analyzing the snapshots of the systems, the lateral diffusion coefficient, $D_{\parallel}$, and 
the lateral radial distribution function, $g_{\parallel}(r)$. To define the nanoparticles
in the same aggregate we defined a minimal distance equals to $r_{min} = 1.2$. If 
the distance between one monomer of one dimer and a monomer of a distinct dimer is smaller
than $r_{min}$ then both dimers belong to the same cluster.

\section{Results and Discussion}
\label{Results}

 \begin{figure}[!h]
  \begin{center}
  \includegraphics[width=8cm]{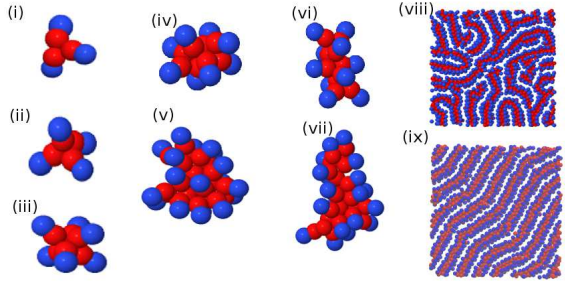}
  \end{center}
  \caption{Aggregates observed in our simulations: (i) trimeric cluster with $n_c = 3 $, (ii) tetrahedral cluster
  with $n_c = 4 $,  (iii) hexahedral cluster with $n_c = 6 $, (iv) spherical cluster with $n_c = 8 $, (v) spherical cluster with $n_c =19 $,
  (vi) elongated cluster with $n_c = 10$, (vii) elongated cluster with $n_c = 20$, (viii)
  disoriented rippled lamellae and (ix) oriented rippled lamellae. Blue particles are the A monomers and red the B monomers. 
  }
  \label{fig2}
  \end{figure}

  We start our discussion showing the distinct micelles observed for the confined system. Regarding
  the temperature and density, the Janus dimers aggregate in clusters with distinct number of nanoparticles
  per cluster, $n_c$. At lower densities, trimeric clusters, with $n_c = 3$
  nanoparticles in each aggregate and tetrahedral clusters, with $n_c = 4$, are more common.
  Increasing the density, hexahedral clusters ($n_c = 6$) are observed, as well spherical and elongated
  micelles with distinct $n_c$. The shape of each aggregate is shown in figure~\ref{fig2}. For densities
  up to a threshold a coexistence of two or three of these micelles was observed. Above the threshold,
  one single rippled lamellae cluster with all the nanoparticles is observed. This lamellae phase can have a 
  disoriented structure, as shown in figure~\ref{fig2}(viii), or an oriented structure, figure~\ref{fig2}(ix).

  \begin{figure}[!h]
  \begin{center}
  \includegraphics[width=8cm]{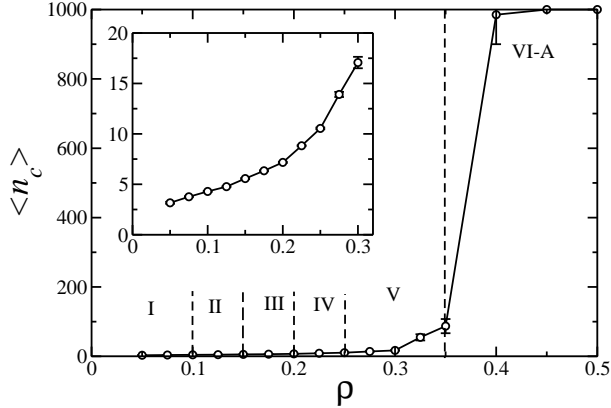}
  \end{center}
  \caption{Mean number of dimers in each cluster, $<n_c>$, as function of density for $T=0.10$. Inset: zoom 
  in the region $\rho < 0.30$. Errors bars smaller than the data point are not shown.}
  \label{fig3}
  \end{figure}
  
\begin{figure}[!h]
  \begin{center}
  \includegraphics[width=8cm]{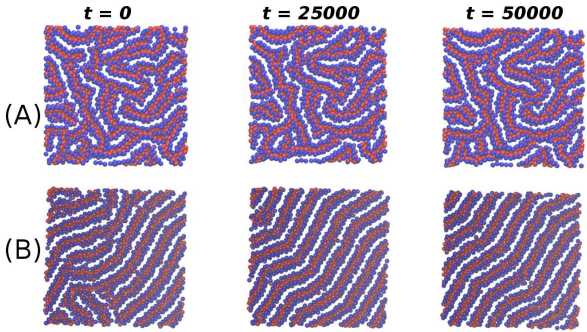}
  \end{center}
  \caption{Lamellar phase for the density $\rho = 0.40$ at temperatures (A) $T = 0.10$ and (B) $T = 0.25$ at three
  distinct times: end of equilibration time ($t = 0$), half of the production time ($t = 25000$) and end of simulation
  ($t=50000$). }
  \label{fig4}
  \end{figure}

  The region of micelles coexistence and the threshold to the lamellae phase depend on the temperature, so lets take the example of $T=0.10$.
  In figure~\ref{fig3}, we show the mean number of dimers in each cluster, $<n_c>$, as function of  the system density.
  We can establish a relation between $<n_c>$ and the type of aggregates. For $0.05 < \rho \leq 0.10$, 
  $<n_c>$ varies between 4 and 5. Analyzing the system snapshots, for these densities we can see a coexistence
  of trimeric, tetrahedral and hexahedral clusters. As $\rho$ increases, more hexahedral aggregates
  and less trimeric aggregates are observed in the solution. This region, with the coexistence of trimeric,
  tetrahedral and hexahedral clusters was labeled as region I in the figure~\ref{fig3}. For the densities
  inside the region II we observe a mixture of tetrahedral and hexahedral clusters, in region III
  hexahedral and small spherical clusters and in region IV a coexistence of small spherical and elongated clusters.
  All these previous aggregates where observed in the bulk simulation, with a similar $<n_c>$~\cite{BoK15c}.
  However, the confinement frustrates the self-assembly as the density increases. Therefore, if for lower densities
  the system aggregates in the same micelles observed for the bulk case, for higher densities new kinds
  of self-assembled aggregates should be induced by the confinement. In this way, the spherical and elongated
  micelles can have a higher $<n_c>$ compared to the bulk case. Basically, the limited space induced by the 
  confinement leads two or more smaller micelles to merge in a large cluster. This region, with spherical 
  and elongated micelles formed by more nanoparticles than the observed in bulk was labeled region V.
  The size of the micelles grows continually up to $\rho = 0.35$ for $T=0.10$, with approximately 10 aggregates with 100
  nanoparticles each, as shown in figure~\ref{fig3}, and above this threshold all the particles aggregate
  in a lamellae structure, region VI-A of figure~\ref{fig3}.
  For the temperature $T=0.10$ and $\rho > 0.40$, the rippled lamellae is disoriented, without a 
  preferable direction, as we shown in figure~\ref{fig2}(viii).
  However, for some values of temperatures and densities, as $T = 0.25$ and $\rho = 0.375$ to $\rho = 0.425$, the lamellar phase
  has a directional ordering, as we shown in figure~\ref{fig2}(ix). The region where we observe the oriented
  rippled lamellae was labeled VI-B. These lamellar structures, usually observed in amphiphilic molecules
  as in Janus particles~\cite{Khan95, Beltran12, Preisler16},
  where not observed in bulk for our model of Janus nanoparticles. Therefore, this new structure is induced by
  the geometrical confinement. While in bulk the particles do not have any geometrical restriction, remaining in
  the micellae phase when the density is up to $\rho = 0.50$~\cite{BoK15c}, the confinement, 
  associated to a high density, leads the dumbbells to aggregate in the lammelar cluster. 
  The time evolution of the distinct lamellar phases is shown in figure~\ref{fig4}. For lower temperatures,
  the entropic contribution to the free energy is not sufficient to change the initial configuration and the
  lamellae structure do not change with time, remaining disoriented, as shown in figure~\ref{fig4}(A). However, for higher temperatures,
  the initially disordered configuration changes to the oriented rippled structure, as we can see in figure~\ref{fig4}(B).
  This lamellar phase with a preferable orientation is characteristic 
  of dumbbells systems with one or two monomer that interacts by a two length scale potential~\cite{Ol10, Bordin16a},
  but at lower temperatures it is frustrated by the confinement.

    \begin{figure}[!h]
  \begin{center}
  \includegraphics[width=8cm]{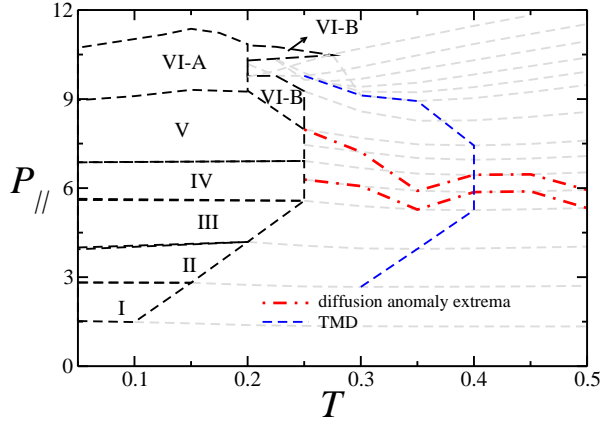}
  \end{center}
  \caption{$P_{||}\times T$ phase diagram for the confined Janus dimer system. The distinct micellae and lamellae regions
  discussed in the previous figures are indicated. The dashed gray lines are the isochores. The density
  anomalous regions is defined by the TMD line, and the diffusion anomaly region by the lines of maxima and minima 
  diffusion.}
  \label{fig5}
  \end{figure}
  
\begin{figure}[!h]
  \begin{center}
  \includegraphics[width=8cm]{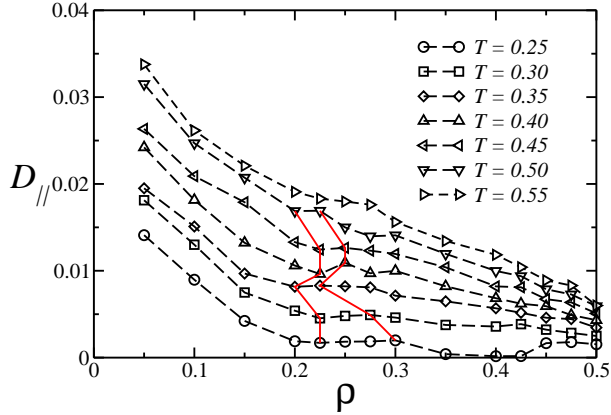}
  \end{center}
  \caption{Parallel diffusion $D_{||}$ as function of density $\rho$ for different temperatures. 
  The red lines indicate the minima and maxima in the diffusion coefficient.}
  \label{fig6}
  \end{figure}

\begin{figure}[!h]
  \begin{center}
  \includegraphics[width=8cm]{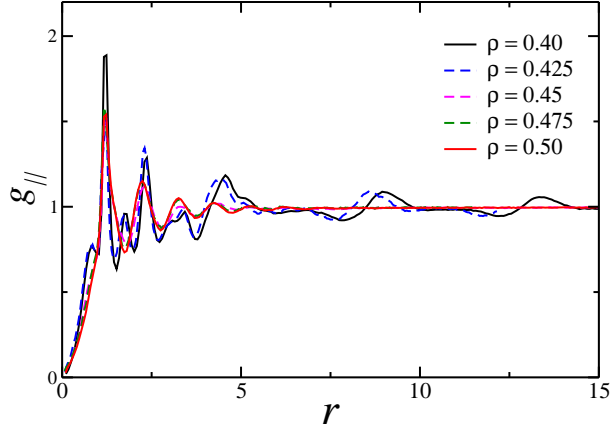}
  \end{center}
  \caption{Lateral radial distribution function (LRDF) $g_{||}(r)$ for $T = 0.25$, showing the melting induced by the density increase.}
  \label{fig7}
  \end{figure}

  In the figure~\ref{fig5} we show the qualitative $P_{||}\times T$ phase diagram for the confined system. The distinct micellae and lamellae
  regions are indicated.   As discussed previously, new self-assembled structures are induced by the confinement. However,
  the aggregation region does not shift to higher or lower temperatures.
  Curiously, at the temperatures were we observe the oriented lamellae structure, for densities above $\rho = 0.45$
  the system have not a well defined micellar structure, and an amorphous phase is observed. This phase is a fluid
  with small diffusion. Taking the isotherm $T = 0.25$ as reference, the pressure drops when
  $\rho > 0.45$ and the system shows a reentrant fluid region. 
  The plot of parallel diffusion coefficient as function of density for some values of $T$ is shown in figure~\ref{fig6}.
  As we can see that for $T = 0.25$ the system have $D_{\parallel} \approx 0$ for $\rho = 0.375$, 0.40 and 0.425, 
  and $D_{\parallel}$ grows when $\rho \ge 0.45$, indicating a melting. The LRDF, showed in figure~\ref{fig7}, also
  shows a decrease in the system structure, from the well structured lamellar phase to the fluid phase.
  This melting induced by the increase of density was already observed in colloidal glasses systems~\cite{Berthier10, Cos13, Everts16}, 
  but was not observed in our bulk system.

  The fluid phase also shows interesting properties, distinct from the bulk case.
  The first one is the density anomaly. As we can see, the isochores in figure~\ref{fig5} show a minimum.
  The dashed blue line, the so called TMD line, connects these minimum points. The density anomaly was observed for pure
  anomalous monomer and dimers (AA dimers)~\cite{Ol06a, Ol10}, however was not present in the bulk Janus system~\cite{BoK15c}.
  The reentrant region, that occurs at higher densities, is located where the TMD line ends
  and splits the lammelar phase VI-B in two regions.

  The bulk system has only diffusion anomaly, and it is preserved in the confined case. 
  However, the diffusion extrema line for the confined system,
  defined by the red dot-dashed red line in the phase diagram, figure~\ref{fig5},
  have a temperature range larger than in bulk.
  Studies for confined anomalous fluids indicate that
  the TMD moves to lower temperatures for solvophobic confinement and higher temperatures for
  solvophilic confinement~\cite{Krott13, Krott14, cas09}.  For the bulk case, our hypothesis 
  was that the TMD line was absorbed by the micellae region. 
  Hence, since our confinement is solvophobic (the WCA 
  potential is purely repulsive), is surprising that the TMD appears, moving to higher temperatures. 
  This leads to the question: why the anomalies lines are shifted to higher temperatures?

\begin{figure}[!h]
  \begin{center}
  \includegraphics[width=8cm]{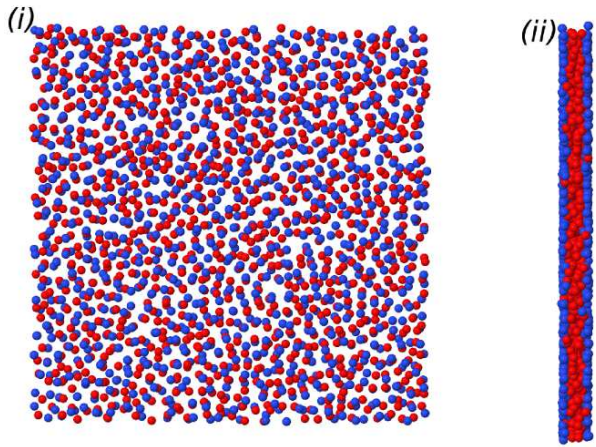}
  \end{center}
  \caption{Frontal (i) and side view (ii) of the system in the fluid region at $T = 0.40$ and $\rho = 0.30$.}
  \label{fig8}
  \end{figure}

  Gavazzoni and co-authors~\cite{Ga14} showed that the anisotropy 
  can shift the solid-fluid phase boundary of dumbbells systems made only by A monomers. 
  More than this, the work argues that the kinetic energy and, therefore, the temperature, has two contribution:
  the translational temperature and the non-translational temperature. Hence, the contribution
  from the dimer rotations to the kinetic energy plays a significant role in this system behavior.
  In bulk, our Janus dumbbell can rotate freely in any direction - the only limitation are collisions with
  others dimers. However, the confinement imposes a constriction to the system. In our strongly confined 
  system, with $L_z=4.0$ and $\lambda = 0.8$, not only the translation in $z$-direction is limited, 
  but the combination of confinement with the competing interactions of Janus systems lead to a interesting
  phenomena. As the figure~\ref{fig6} shows, the diffusion coefficient is distinct from zero -- so, the 
  particles are moving in the parallel direction. In the figure~\ref{fig8}(i) we show a frontal snapshot
  of the nanoparticles arrangement. As we can see, the particles are disordered - as expected for fluids.
  Notwithstanding, the side snapshot shows that in the $z$-direction the particles have a preferential 
  position: the attractive B particles stay at the center, and the repulsive A particles are near
  the wall. Then, due the confinement and the Janus characteristics,
  the dimers are translating in the $xy$-plane, but without rotation.
  This places the A monomers side by side in the $xy$-plane, with a internal layer of attractive B monomers.
  This internal layer will act similar to a solvophilic wall, with the B monomers pulling one another.
  As consequence, the behavior is similar to water in hydrophilic confinement, and the anomalous region
  shifts to higher temperatures.

%
%

\section{Conclusion}
\label{Conclu}

 We report the study of confined Janus nanoparticles. This system has special interest in the design of
 new material using the confinement to control the self-assembled structures. We have found a rich variety
 of aggregates and micelles, including large micelles not observed in the bulk system. 
 As well, two lamellae phases, with distinct orientations, where induced by the confinement.
 The oriented lamellae phase region in the $P_{\parallel}T$ phase diagram is splitted 
 by a reentrant fluid phase. This melting, induced by increasing the density, was not 
 present in the bulk system as well.

 Another feature that was not observed in the bulk system is the density anomaly. 
 The combination of confinement effects with the competing interactions of Janus dimers
 shifts the TMD line to higher temperatures, rising the anomaly
 that was hide inside the aggregation region for the bulk case. As well, the diffusion
 anomaly region increases, reaching higher temperatures. 
 
 Our results show that materials composed by the association of a  monomer which can be modeled by a 
 two length scale potential and a standard LJ monomer have an interesting
 and peculiar behavior.

\section{Acknowledgments}

The authors are grateful to  Marcia C. Barbosa from Universidade Federal do Rio Grande do Sul for valuable
and critical discussion. JRB thanks the Brazilian agency CNPq for the financial support.

\providecommand*{\mcitethebibliography}{\thebibliography}
\csname @ifundefined\endcsname{endmcitethebibliography}
{\let\endmcitethebibliography\endthebibliography}{}

  \end{document}